\documentclass[twoside,fleqn]{ActaStyle}
\usepackage[dvips]{graphics}
\usepackage{times,cite}
\usepackage{xspace}
\usepackage{graphicx}
    \def\authorheading{H. Caines et al.}
    \def\shorttitle{Effects of Volume on Strangeness Production.}

\newcommand{\xim}{$\Xi^-$}
\newcommand{\sqrts}{$\sqrt{s_{_{NN}}}$}
\newcommand{\pT}{$p_{T}$\xspace}
\newcommand{\Npart}{$\langle N_{part}\rangle$\xspace}
\newcommand{\Nbin}{$\langle N_{bin}\rangle$\xspace}
\newcommand{\pp}{{\it p-p}\xspace}

\begin{document}
    \pagerange{1}{4}   

    \title{The Effects of Varying the Correlation Volume on Strangeness
Production in High Energy Collisions}

    \author{H.~Caines\email{helen.caines@yale.edu} for the STAR Collaboration}
              {WNSL, Physics Department, Yale University, New Haven, CT 06520, U.S.A.}

    \abstract{Preliminary results on strange particle production versus
collision centrality are presented. STAR measurements from \sqrts =
200 GeV heavy-ion and \pp collisions are compared to SPS
measurements. A systematic study of strange particle production is
presented with the aim of establishing how the correlation volume of
the produced source affects the scale of strange particle creation,
including that of the multi-strange baryons. A linear increase of
strangeness production with volume has been suggested by thermal
models as an indication that the collision region has reached
sufficient size such that small volume effects can be neglected.
Analysis of preliminary results from STAR show that, using the
assumption that the number of participants is linearly correlated
with the volume, no such regime was obtained. This suggests that the
correlation volume ''seen" by strange quarks is not merely that of
the initial overlap.}
       \pacs{...}

\noindent  By studying the spectra and yields of the strange
particles produced in heavy-ion collisions and comparing them to
those resulting from elementary \pp collisions, in which a QGP phase
is not expected, we gain insight into the properties of the medium.
While we cannot measure a state of free quarks and gluons directly,
indirect methods can be used to provide evidence that such a phase
did exist and whether the state was in chemical and/or thermal
equilibrium. It is predicted that strangeness is suppressed in \pp
due to a lack of phase space~\cite{Redlich}. As the collision volume
increases, this suppression reduces. The relevant volume is believed
to be linearly proportional to the number of participants in the
collision, \Npart. Calculations also show that the \pp suppression
increases with the strangeness content of the particle and decreases
with increased collision energy. The scale of the suppression is
 very sensitive to the temperatures of the
produced systems. For large systems this phase space suppression is
removed and strangeness is produced freely, resulting in yields that
are proportional to the volume. It is then appropriate to apply the
"Grand Canonical" approach, where quantum numbers need only be
conserved on average, to calculate the yields. Experimentally the
scale of the suppression is measured via a so-called enhancement
factor:
\begin{equation}\label{Eqn:Enhance}
    E(i) = (Yield(i)_{AA}/N_{part} )/ (Yield(i)_{pp}/N_{part})
\end{equation}

\noindent Detailed predictions of the phase space suppression
effects do not yet exist for 200 GeV collisions. However, for
central events using T=177 MeV and $\mu_{b}$=29 MeV enhancements of
E(\xim)=3.05 and E($\Lambda$)=1.44 are calculated. For T=170 MeV and
$\mu_{b}$=29 MeV, E(\xim)=4.5 and E($\Lambda$)=1.77~\cite{Redlich2}.

~

\noindent We present here preliminary data taken with the STAR
experiment from Au-Au and \pp collisions at \sqrts= 200 GeV. The \pT
spectra and yields have been analyzed as a function of centrality
and a Glauber model is used to calculate \Npart and \Nbin for each
centrality bin~\cite{Glauber}.

\begin{figure}[htbp]
 \begin{minipage}{0.5\linewidth}
    \begin{center}
   \includegraphics[width=0.9\linewidth]{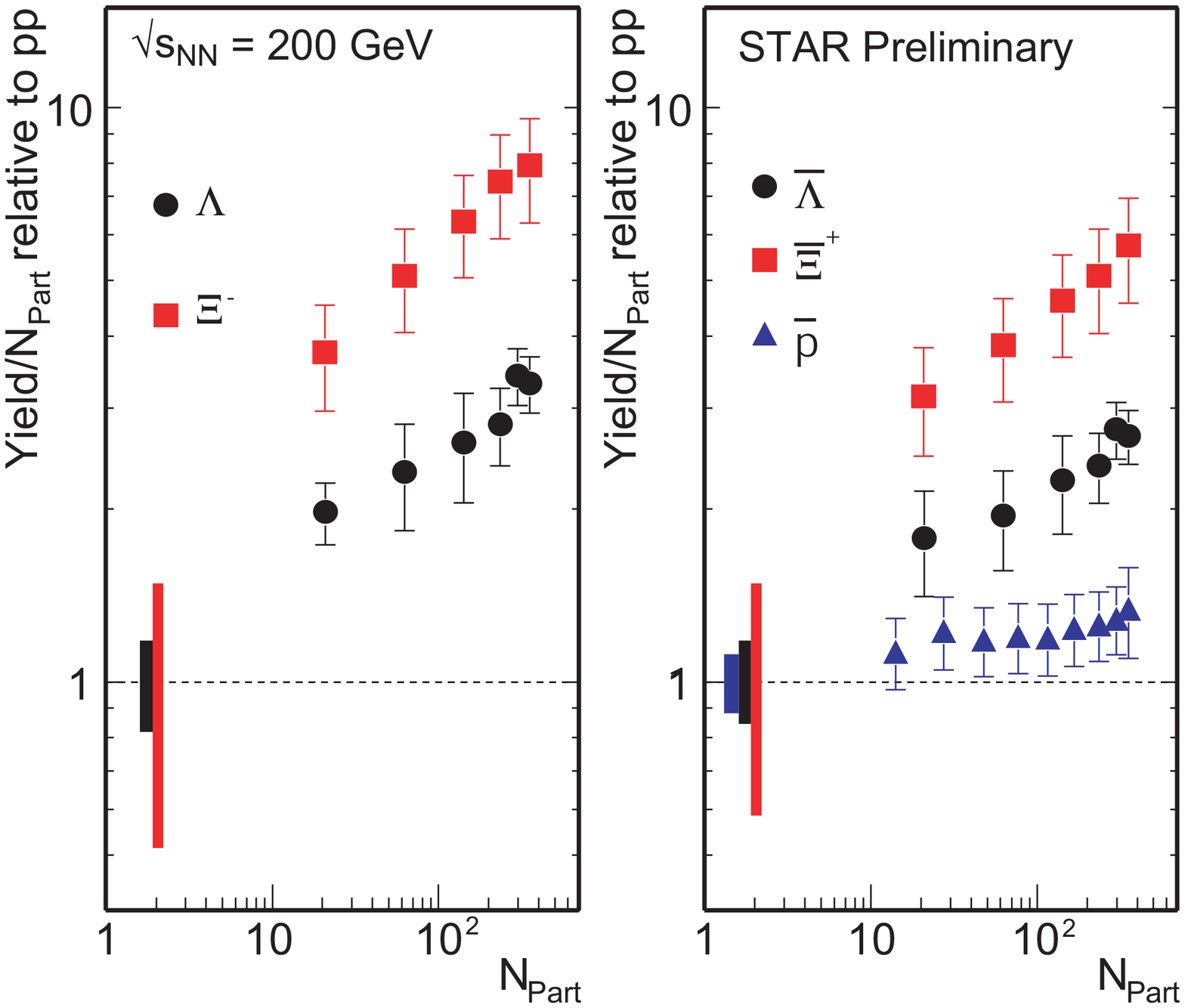}\\
   \vspace{-0.3cm}
\caption{ Preliminary measured enhancements for \sqrts=200 GeV Au-Au
collisions. The uncertainties shown are a combination of statistical
and systematic.} \label{Fig:Enhance}
\end{center}
 \end{minipage}
 \hspace{0.21cm}
\begin{minipage}{0.48\linewidth}
 \begin{center}
  \includegraphics[width=\textwidth]{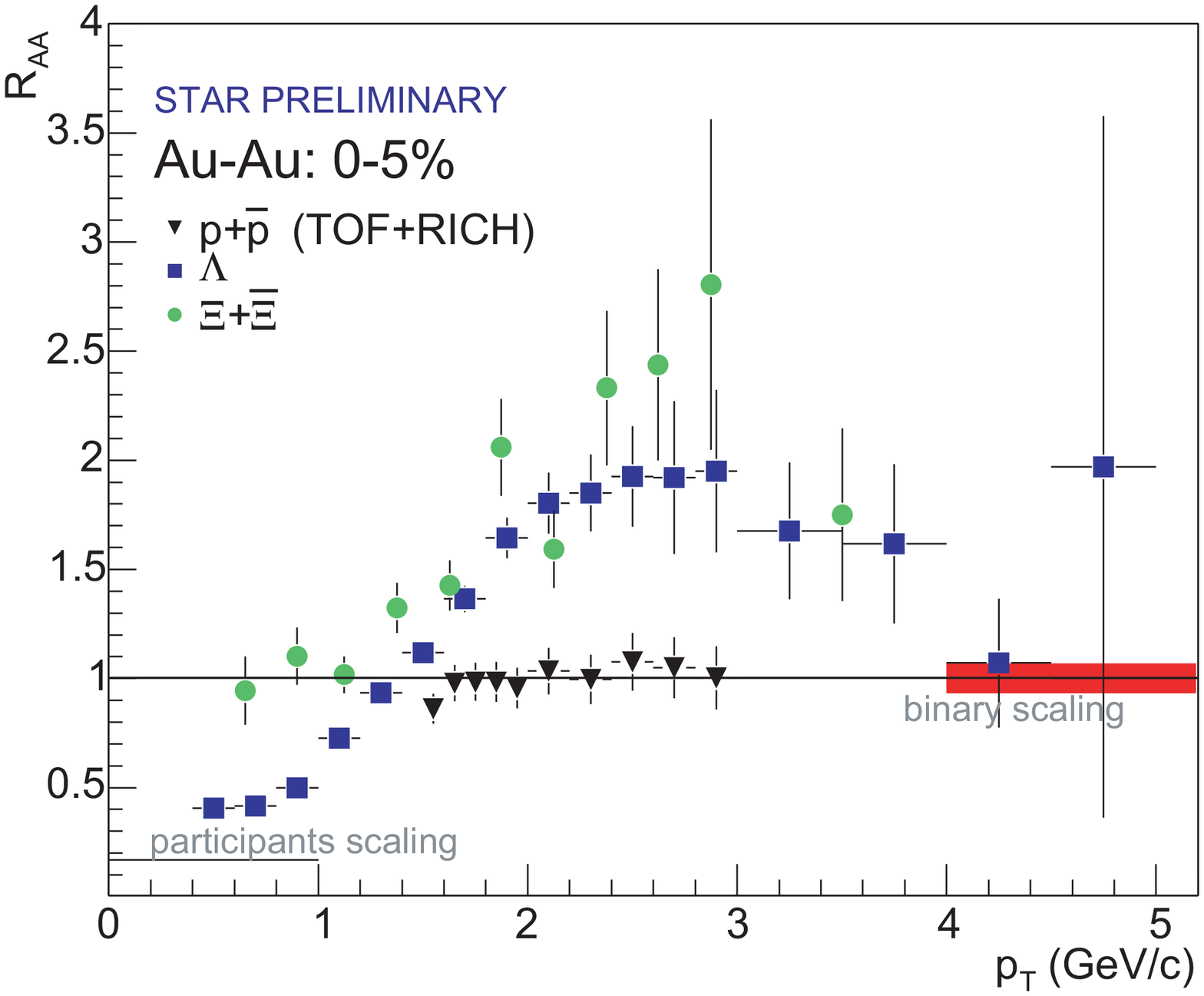}

\caption{ Preliminary measurements of the nuclear modification
factor for baryons. The dotted curve represents the charge hadron
measurement.} \label{Fig:Raa}
   \end{center}
   \end{minipage}
\end{figure}

~

\noindent Fig.~\ref{Fig:Enhance} shows the preliminary measured
enhancement factors from STAR. The $\overline{p}$ show a linear
scaling with \Npart and almost no enhancement relative to \pp.
Meanwhile,  the $\Lambda$ and the $\Xi$ exhibit significant
enhancements, even for the most peripheral data. The expected
ordering with strangeness content is also observed, i.e. the $\Xi$
are more enhanced than the $\Lambda$. The particles and their
anti-particles show similar enhancements - reflecting the near-zero
net baryon number. However, neither the $\Lambda$ nor $\Xi$ show a
linear scaling with \Npart, suggesting that the Grand Canonical
regime is not reached even in central events. In contrast,
calculations of $\gamma_{s}$, the strangeness saturation factor,
using a thermal model show that it approaches unity for the more
central bins from fits to STAR data\cite{StarSpectra}. These fits
therefore support the use of the Grand Canonical approach and
suggest that  the strangeness yields should scale with the collision
volume. As these results are contradictory, we conclude that the
relevant volume for strangeness production is not linearly
proportional to \Npart and hence not purely related to that of the
initial collision overlap. This idea allows for the Grand Canonical
regime to have been reached, as suggested by thermal model fits
together with a lack of proportionality of the yields with \Npart.

\begin{figure}[htbp]
 \begin{minipage}{0.49\linewidth}
 \vspace{-0.2cm}
    \begin{center}
\includegraphics[width=0.9\textwidth]{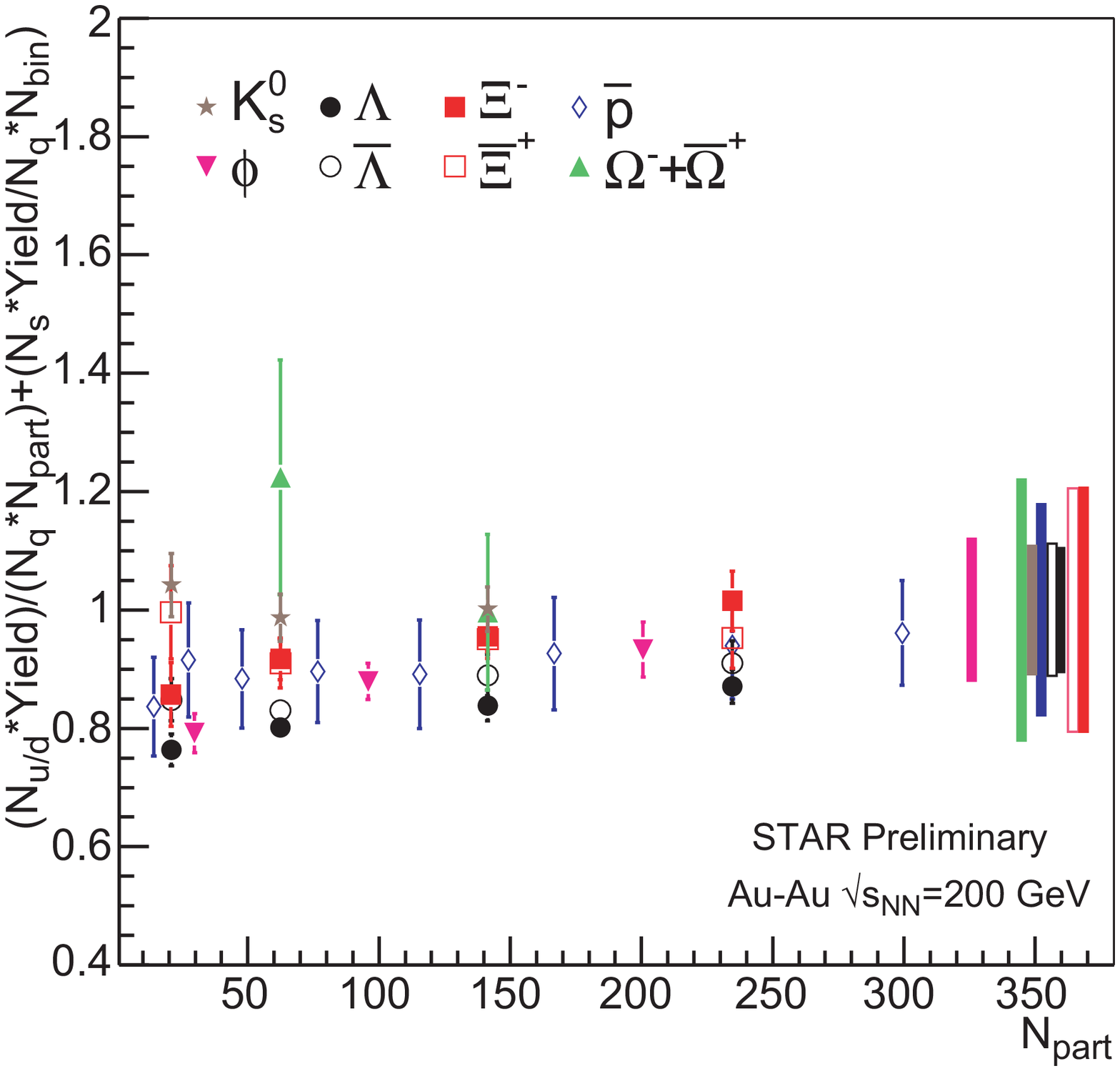}
 \caption{ Preliminary Au-Au data scaled as defined by
Eqn.~\ref{Eqn:Scale}, except the $\phi$ which is scaled by \Npart.
All data are normalized to the most central bin. } \label{Fig:Scale}
\end{center}
 \end{minipage}
\hspace{0.2cm}
\begin{minipage}{0.49\linewidth}
 \begin{center}
\includegraphics[width=\textwidth]{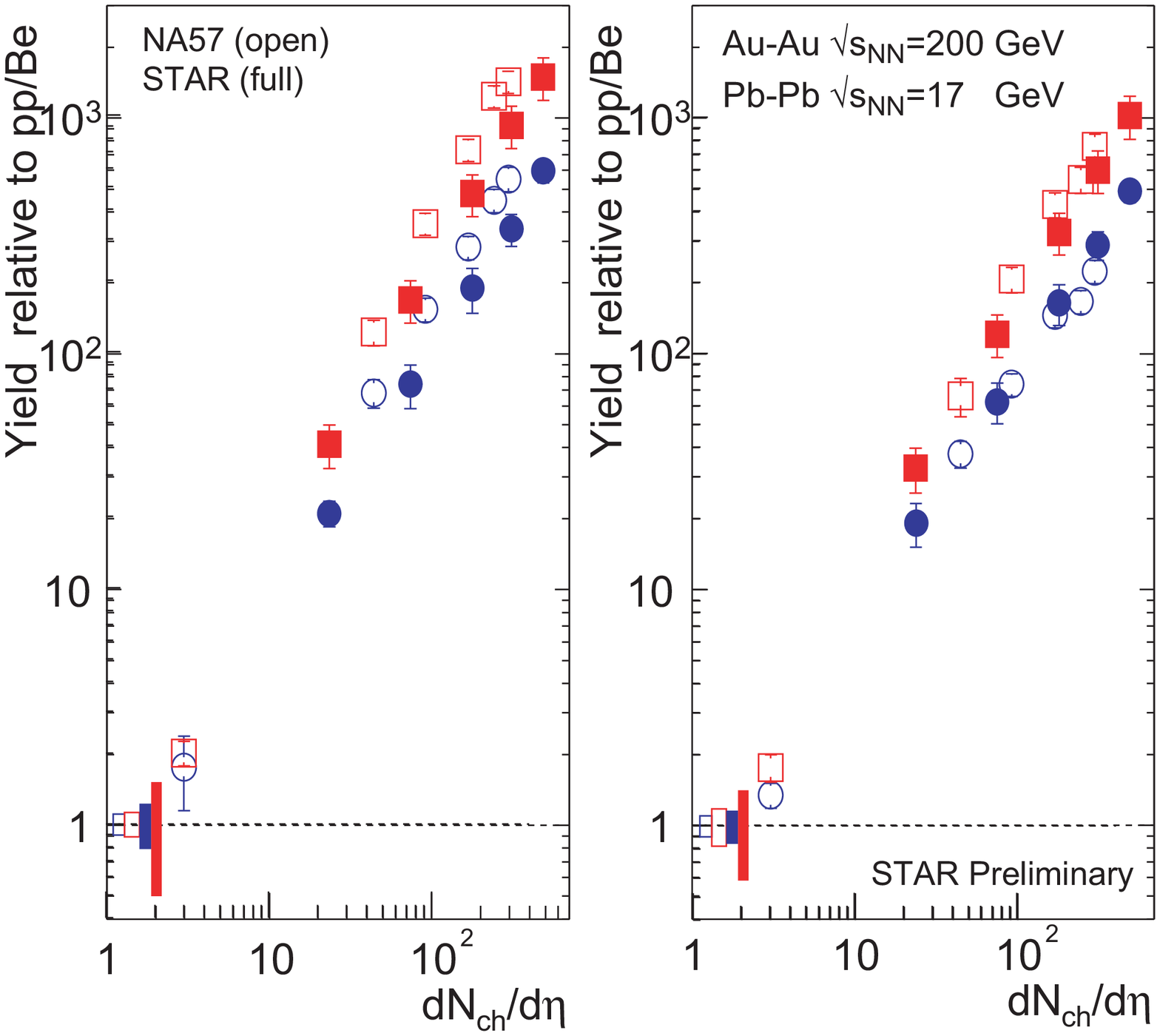}
\caption{Preliminary RHIC and SPS strange baryon yields relative to
the measured \pp yields as a function of dN$_{ch}$/d$\eta$. Open
symbols are for the SPS data and closed symbols from STAR.}
\label{Fig:EnhanceNch}
   \end{center}
   \end{minipage}
\end{figure}

~

\noindent In order to explore how strangeness production in the
central Au-Au data compares to \pp as a function of \pT, we
calculate the nuclear modification factor, R$_{AA}$, for baryons.
The preliminary results are shown in Fig.~\ref{Fig:Raa}. Previously
reported measurements of R$_{CP}$~ (the ratio of
central/peripheral), show that all baryons follow the same curve and
are suppressed at high \pT, the result of the modification of jets
by the medium\cite{StarHighPt}. Our measured R$_{AA}$, however,
reveals a striking dependency on the strangeness content. This can
be explained by considering that R$_{CP}$ uses peripheral events to
determine the baseline spectra and R$_{AA}$ uses \pp data.
Fig.~\ref{Fig:Enhance} has already shown that the peripheral Au-Au
yields are enhanced compared to scaled \pp. Thus, R$_{AA}$  reveals
a combined effect of suppressed \pp production, due to phase space
effects, and suppressed Au-Au production, due to "jet quenching".
The R$_{CP}$ measure only reveals the jet quenching. What is
unexpected is that the "soft" physics phase space suppression in \pp
appears to dominate the R$_{AA}$ ratio out to \pT $>$ 3 GeV/c.

~

\noindent Since the Au-Au strangeness data appear to show that
\Npart, or the initial nuclear overlap, is not the appropriate
scaling volume for strangeness production we take a closer look at
the centrality dependence.  Our goal is to try to determine if there
is a measurable quantity that is proportional to the strangeness
production volume – in which case the yields as a function of
centrality will scale linearly.  Fig.~\ref{Fig:Enhance} shows that
the anti-proton alone scales with \Npart. A closer look at the
strange baryons reveals that as strange quarks are added to the
particle, the scaled yield deviates more strongly from a linear
dependence on \Npart and in fact comes closer to \Nbin scaling. This
suggests that the strange quarks have a large \Nbin contribution
whereas  the light u/d quarks scale with \Npart. This idea is
supported by R$_{AA}$ (Fig.~\ref{Fig:Raa}) which shows that the
strange baryons scale significantly above that of simple \Nbin
scaling at intermediate \pT. We therefore consider a normalization
that is dependent on the quark content of the particle:

\begin{equation}\label{Eqn:Scale}
   C_{scaling} = N_{light}*N_{part}/N_{q} + N_{s}*N_{bin}/N_{q}
\end{equation}

\noindent where N$_{q}$ is the number of quarks in the particle,
N$_{light}$ is the number of light (u and d) quarks, and N$_{s}$ is
the number of strange quarks. While this scaling is mostly
successful, the $\phi$, which according to  Eqn.~\ref{Eqn:Scale}
should scale with \Nbin, appears to scale with  \Npart. This \Npart
scaling is used in Fig.~\ref{Fig:Scale}, while all other particles
are scaled by Eqn.~\ref{Eqn:Scale}. It can be seen that the scaling
is good to $\sim$20$\%$ level.

~

\noindent The PHOBOS collaboration has reported a strong correlation between
N$_{ch}$/d$\eta$ and \Npart~\cite{Phobos}. A two component fit

\begin{equation}\label{Eqn:Nch}
    N_{ch}/d \eta= n_{pp}((1-x)\langle N_{part} /2 \rangle + x \langle N_{bin} /2 \rangle
\end{equation}
was used to fit the data. Where $n_{pp}$ = 2.29 (1.27) at \sqrts=
200 (19.7) GeV  and is the mid-rapidity charged particle yield in
$\overline{p}$(p)-p collisions. $x$ represents the contribution to
the yield from hard processes. Independent fits to the data using
Eqn.~\ref{Eqn:Nch}, leaving $x$ as a free parameter, determined that
the 200 GeV and 19.6 GeV Au-Au data resulted in the same value of
$x$. A simultaneous fit gave $x = 0.13\pm0.01\pm0.05$. This suggests
that the contributions from hard processes are independent of
collision energy, a result not predicted by pQCD calculations.
Building on the idea from PHOBOS,  M. Lisa et al.~\cite{HBT} have
recently reported that when the  measured HBT radii are plotted
versus dN$_{ch}$/d$\eta^{1/3}$, a linear scaling is observed. Their
reasoning is that HBT correlation measurements are related to the
source size and therefore they must have some dependence on the
initial geometrical overlap, i.e. $\langle N_{part}\rangle^{1/3}$.
This, however, does not give a perfect scaling, which could be due
to the fact that $\langle N_{part}\rangle^{1/3}$ relates to the
original overlap volume and HBT measurements resolve the source's
final state. Since dN$_{ch}$/d$\eta$ is closely related to final
state density, this variable, or rather dN$_{ch}$/d$\eta^{1/3}$
which gives an approximate length scale, should result in a stronger
correlation, and indeed it does. Fig.~\ref{Fig:EnhanceNch} shows the
strangeness yields scaled by the \pp(Be) data from STAR and NA57 at
the SPS~\cite{NA57} as a function of dN$_{ch}$/d$\eta$, estimated
using Eqn.~\ref{Eqn:Nch} and the calculated values of \Npart and
\Nbin. This results in an apparent energy independent  scaling for
the strange anti-baryons which are linear in dN$_{ch}$/d$\eta$.
Eqn.~\ref{Eqn:Nch} shows that dN$_{ch}$/d$\eta$ combines
contributions from \Npart and \Nbin which seems to be necessary to
describe  strangeness production in A-A collisions.

~

\noindent In summary, strange baryon production is not trivially
related to the number of participants in the collision and the
number of binary collisions seems to have a significant impact on
the \pT integrated yields. The measured R$_{AA}$ of these baryons
suggests that the strong effect of the phase space suppression
extends out to high \pT. Meanwhile, three independent observations
appear to provide evidence that dN$_{ch}$/d$\eta$, which is closely
related to the entropy of the system, is the underlying drive for
many of the global observables measured in heavy-ion collisions.

\end{document}